\documentclass{article}
\usepackage{amsmath,amssymb,amsfonts,amsthm}
\usepackage{color}
\usepackage{graphicx}
\usepackage{wrapfig}

\begin{document}

\flushbottom

\numberwithin{equation}{section}

\renewcommand{\figurename}{Fig.}
\def\refname{References}
\def\proofname{Proof}

\newtheorem{theorem}{Theorem}
\newtheorem{propos}{Proposition}
\newtheorem{remark}{Remark}

\def\tens#1{\ensuremath{\mathsf{#1}}}

\if@mathematic
   \def\vec#1{\ensuremath{\mathchoice
                     {\mbox{\boldmath$\displaystyle\mathbf{#1}$}}
                     {\mbox{\boldmath$\textstyle\mathbf{#1}$}}
                     {\mbox{\boldmath$\scriptstyle\mathbf{#1}$}}
                     {\mbox{\boldmath$\scriptscriptstyle\mathbf{#1}$}}}}
\else
   \def\vec#1{\ensuremath{\mathchoice
                     {\mbox{\boldmath$\displaystyle\mathbf{#1}$}}
                     {\mbox{\boldmath$\textstyle\mathbf{#1}$}}
                     {\mbox{\boldmath$\scriptstyle\mathbf{#1}$}}
                     {\mbox{\boldmath$\scriptscriptstyle\mathbf{#1}$}}}}
\fi

\newcommand{\diag}{\mathop{\rm diag}\nolimits}
\newcommand{\const}{\rm const}
\def\div{\operatorname{div}}

\begin{center}
{\Large\bf The Chaplygin sleigh with parametric excitation: chaotic dynamics and nonholonomic acceleration}

\bigskip

{\large\bf Ivan~A.\,Bizyaev$^1$,
Alexey~V.\,Borisov$^2$,
Ivan~S.\,Mamaev$^3$\\}
\end{center}

\begin{quote}
\begin{small}
\noindent
$^{1}$ Moscow Institute of Physics and Technology, \\
Institutskii per. 9, Dolgoprudnyi, 141700 Russia \\

$^{2}$ Udmurt State University, \\
ul. Universitetskaya 1, Izhevsk, 426034 Russia \\

$^{3}$ Izhevsk State Technical University, \\
ul. Studencheskaya 7, Izhevsk, 426069 Russia

$^1$ E-mail: bizaev\_90@mail.ru \\
$^2$ E-mail: borisov@rcd.ru \\
$^3$ E-mail: mamaev@rcd.ru
\end{small}

\bigskip
\bigskip

\begin{small}
\textbf{Abstract.}
This paper is concerned with the Chaplygin sleigh with time-varying mass distribution (parametric excitation).
The focus is on the case where excitation is induced by a material point that executes periodic oscillations in a
direction transverse to the plane of the knife edge of the sleigh. In this case, the problem reduces to investigating
a reduced system of two first-order equations with periodic coefficients, which is similar to various nonlinear parametric
oscillators.
Depending on the parameters in the reduced system, one can observe different types of motion, including those accompanied by
strange attractors leading to a chaotic (diffusion) trajectory of the sleigh on the plane. The problem of unbounded acceleration
(an analog of Fermi acceleration) of the sleigh is examined in detail.
It is shown that such an acceleration arises due to the position of the
moving point relative to the line of action of the nonholonomic constraint and the center of mass of the platform.
Various special cases of existence of tensor invariants are found.
\smallskip


\smallskip

\end{small}
\end{quote}

\footnotetext[0]{The work of A.\,V.\,Borisov (Sections~\ref{biz-sec1} and~\ref{biz-sec2}) was carried out within the framework of the state assignment of the Ministry of Education
and Science of Russia (1.2404.2017/4.6). The work of I.\,A.\,Bizyaev (Section~\ref{biz-sec3}) was carried out at MIPT under project 5-100
for state support for leading universities of the Russian Federation. The work of I.\,S.\,Mamaev (Section~\ref{biz-sec4}) was carried out within
the framework of the state assignment of the Ministry of Education and Science of Russia (1.2405.2017/4.6).}

\section*{Introduction}

{\bf 1.} The Chaplygin sleigh on a plane is one of the well-known model systems of nonholonomic mechanics.
According to S.\,A.\,Chaplygin~\cite{bizyaev_b21-3}, the sleigh can be made by attaching a knife edge and two
absolutely smooth legs to a rigid body.
A nonholonomic constraint in this case is generated by the knife edge: the translational velocity at the point of contact of
the knife edge is orthogonal to its plane (that is, to the body-fixed direction).
A similar constraint can also be realized by using a wheel pair instead of the knife edge~\cite{HadamardHamel}.

The free dynamics of the Chaplygin sleigh on a horizontal plane was studied by C.\,Carath\'{e}odo\-ry~\cite{Caratheodory}.
Depending on the position of the center of mass relative to the knife
edge, the sleigh moves in a circle or asymptotically tends to straight-line motion. In the latter case,
the classical scattering problem arises for which the angle of scattering is indicated in~\cite{Fuf}.
It is calculated explicitly, since the free motion of the sleigh is integrable and regular~\cite{Caratheodory}.
The dynamics of the Chaplygin sleigh on an inclined plane is no longer integrable and exhibits random asymptotic behavior
depending on the initial conditions~\cite{BM2}.

The recent paper~\cite{Jung} investigates the motion of the Chaplygin sleigh under the action of random forces, which
simulate a fluctuating continuum. It turns out that in this case the sleigh exhibits intricate behavior, which,
according to the authors, resembles random walks of bacterial cells with some diffusion component.
A similar behavior is exhibited by the sleigh under the action of angular momentum depending on its orientation and in
the presence of viscous friction~\cite{BorisovKuznetsov}. Other generalizations of the problem of the Chaplygin sleigh were
considered in~\cite{bizyaev, bizyaevCylinder, JacobiIntegral}.

{\bf 2.} This paper addresses various aspects of the dynamics of a nonautonomous Chaplygin sleigh (that is, with time-varying
mass distribution).
Special attention is given to the case in which the center of mass of the sleigh periodically changes with time. In
practice, this can be achieved by means of various mechanisms such as eccentrics and sliders placed inside the body.
Such control mechanisms were discussed in the problem of the planar motion of a rigid body in an ideal
fluid~\cite{KozlovPMM2004, Ramod2002, KozlovOn, KozlovRam, KilinVed, Vetchanin}.
The study of the problem at hand, as opposed to the above-mentioned problem, reveals much more new dynamical effects due to the
absence of an additional integral of motion (similar to angular momentum).

In this paper, we study the dynamics of the Chaplygin sleigh with parameters periodically depending on
time. This study is closely related to the control problem. Since the sleigh can be made in the form of
a two-wheeled robot~\cite{HadamardHamel},
this study is of great practical importance, since the regimes arising at fixed values of angular velocities of the eccentrics
can be taken as basic regimes (the so-called gaits), which the body reaches after various maneuvers initiated by the control system.
Problems of controlling the Chaplygin sleigh by displacing the center of mass are addressed in~\cite{Zenkov},
in which attention is given to a maneuver necessary for a transition from motion in a circle to straight-line motion.
As far as we know, a general study of the control of the Chaplygin sleigh in the spirit of the Rashevsky\,--\,Chow theorem
has not been carried out so far. We note that periodic changes in control functions were also considered in optimal
control problems~\cite{Leonard, Murray}.

{\bf 3.} This paper presents a detailed study of the dynamics of a reduced system (which decouples from the complete system
of equations),
which describes the evolution of translational and angular velocities of the sleigh.
The dynamics of the point of contact is defined by quadratures from known solutions of a reduced system.

A reduced system is a system of two first-order equations with periodic coefficients.
However, in contrast to Hamiltonian systems with one and a half degrees of freedom, it has no smooth invariant measure~\cite{BM2}
and can have various (including strange) attractors typical of dissipative systems. In this sense, it is similar to oscillators
with parametric periodic excitation of Duffing and van der Pol type~\cite{Sprott} and to nonlinear Mathieu equations~\cite{Izrailev,Ito}.
However, as noted in many papers, ``nonholonomic dissipation'', which arises due to sign-alternating divergence,
possesses specific features that require an additional study. Starting with~\cite{b21-75}, strange attractors of different
nature~\cite{BorisovTop,Kuznetsov2012,Gonchenko2013} have been observed in nonholonomic systems.

As a rule, the presence of a strange attractor in the reduced system leads to chaotic (or even diffusional)
behavior of the contact point of the knife edge of the sleigh with no explicit directed drift. From the viewpoint of
control, one should either avoid such dynamics or use it based on chaos control methods~\cite{Ott}.

{\bf 4.} For the study of the dynamics of a nonautonomous Chaplygin sleigh, its acceleration poses a more interesting problem.
From the physical point of view, interest in this problem is motivated by the fact that unbounded increase in energy, and hence
unbounded acceleration, is achieved by a mechanism performing small, but regular oscillations.

Such an acceleration cannot be achieved in nonholonomic systems with other control mechanisms~\cite{45} (for example, for a ball
controlled by rotors~\cite{ControlI, ControlII}) and requires constantly increasing the angular velocities of rotation of rotors in such
systems. This often makes them useless in practice. In~\cite{Pivovarova}, the motion of the ball is controlled using
a pendulum-type mechanism, which can lead to an unbounded acceleration.

{\bf 5.} As noted above, the system dealt with in this paper differs from Hamiltonian systems with one and a half degrees of
freedom.
This difference becomes particularly apparent in the case of acceleration.
The Hamiltonian model of acceleration started to be discussed in physical studies in connection with Fermi acceleration
in Ulam's model~\cite{Lichtenberg}, which reduces to investigating an area-preserving two-dimensional Poincar\'{e} map.
As shown numerically and then proved analytically, acceleration in different variations of Ulam's model is impeded by the
existence of an invariant curve at large velocities, which is predicted by KAM theory. In order for
acceleration to become characteristic in nonlinear natural Hamiltonian systems, it is necessary to consider systems with two and
a half degrees of freedom; in this case, acceleration is closely related to Arnold's diffusion.
There are already a number of such systems, in which acceleration is shown numerically\\~\cite{Lenz, Pereira} or
by analytical methods allowing the presence of trajectories with increasing energy to be proved~\cite{Bolotin, Koiller, Turaev}.
An interesting example, which has been intensively discussed recently, is the two-dimensional periodically pulsating
Birkhoff billiard. Depending on the shape of the boundary, which determines the dynamical (stochastic, ergodic or regular) behavior
of  the ``frozen system'', different degrees of increase in the particle's energy are possible when pulsation is introduced.

In nonholonomic mechanics, since there is no continuous invariant measure, acceleration is already typical for small dimensions
and takes place, in particular, in the system under consideration. In this paper, we present explicit solutions
possessing acceleration, and analyze numerically conditions on parameters defining the region of acceleration. We note that
most problems in this direction still remain open; for example, no analysis has been made of the possibility of
acceleration in the entire parameter region.
We mention the recent paper~\cite{Kuznetsov}, in which the acceleration of the Chaplygin sleigh is studied
by the averaging method and the asymptotics of the degree of acceleration depending on time is obtained.

{\bf 6.} We discuss a number of related problems, which can be investigated by the methods presented in this paper.
A hydrodynamical model of the Chaplygin sleigh was proposed in~\cite{Fedorov}. Although it requires additional justification
from the viewpoint of hydrodynamics, it would be interesting to explore a nonautonomous analog of this model.
The same can be said of the problem of a sleigh with a constraint inhomogeneous in velocities, which has been treated recently
in~\cite{BM}.

\begin{figure}[!ht]
\centering
		\includegraphics[totalheight=3cm]{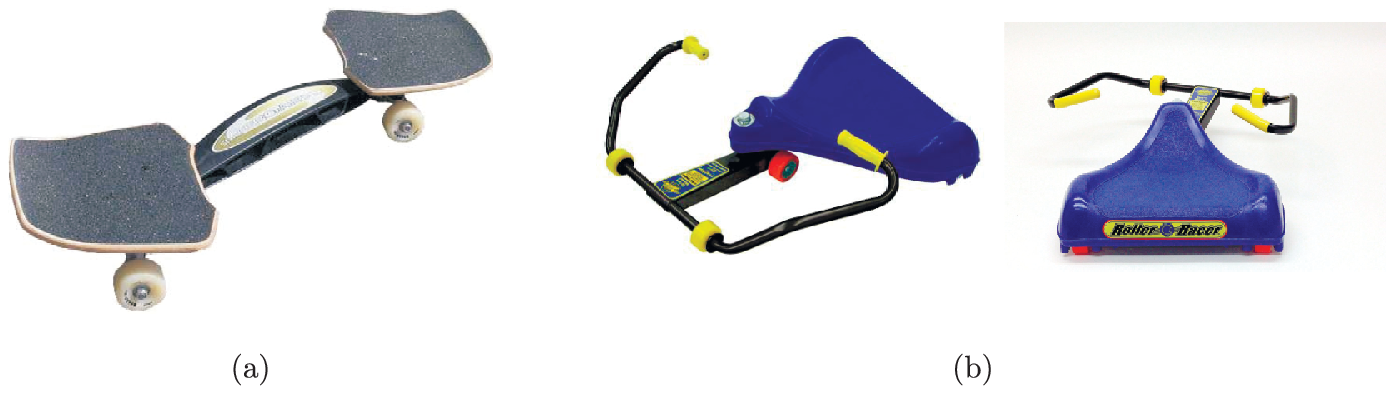}
		\caption{ }
		\label{SR}
\end{figure}

Also of interest is the possibility of acceleration in more complex nonholonomic systems such as  the {\it snakeboard}~\cite{Ostrowskiy}
(see Fig.~\ref{SR}a) and the {\it roller-racer}~\cite{Krishnaprasad}  (see Fig.~\ref{SR}b), which are an immediate generalization
of the sleigh problem. In both systems, the maneuvering of motion is achieved by periodically changing the dynamical parameters.

In describing the {\it snakeboard}, the human body is usually simulated by a balanced rotor.
A special feature of the {\it roller-racer} is that a person sitting on it moves forward by oscillating the handle bar
in a transverse direction from side to side. These actions can be interpreted as periodic oscillations
of the material point in the transverse direction.
We also mention the papers~\cite{KellyAbrajan,KellyBeanie, Tallapragada}, in which the nonholonomic Chaplygin sleigh
is related to hydrodynamical robotic problems.
Nevertheless, such a use of nonholonomic equations for describing the motion of a body in a fluid is not correct~\cite{Biz-A4,Biz-A5}.

\section{Equations of motion}
\label{biz-sec1}

\begin{figure}[!ht]
\centering
		\includegraphics{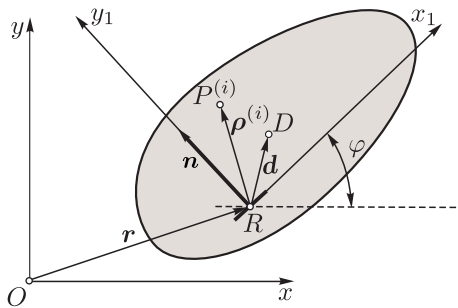}
		\caption{The Chaplygin sleigh.}
		\label{fig01}
\end{figure}

We explore the dynamics of a mechanical multicomponent system with a nonholonomic constraint. The system consists of a platform,
which slides on a horizontal plane like the Chaplygin sleigh~\cite{bizyaev_b21-3}, that is, the body-fixed point $R$ (see
Fig.~\ref{fig01}) cannot slide in some direction $\boldsymbol n$ fixed relative to the platform:
\begin{equation}
\label{eq1}
(\boldsymbol v_R, \boldsymbol n)=0.
\end{equation}
On this platform, $n$ material points $P^{(i)}$, $i=1, \ldots, n$ move according to a given law.
\begin{remark}
This approach can be generalized in a natural way to the case of motion of an arbitrary rigid body deformable by a given law.
\end{remark}

To describe the motion of the system, we define two coordinate systems:
\begin{itemize}
	\item[---]  a {\it fixed} (inertial) coordinate system $Oxy$;
	\item[---]  a {\it moving} coordinate system $Rx_1y_1$ attached to the platform.
\end{itemize}

The position of each point relative to the platform is defined  by the radius vector in the moving coordinate system:
$$
\boldsymbol {\rho}^{(i)}(t)=\Big(\rho^{(i)}_1(t), \rho^{(i)}_2(t)\Big), \quad i=1,\ldots, n.
$$

We specify the position of the platform by the coordinates $(x, y)$ of point $R$ in the fixed coordinate system $Oxy$,
and its orientation by the angle of rotation $\varphi$ (see Fig.~\ref{fig01}). Thus, the configuration space of the
system $\mathcal{Q}=\{\boldsymbol {q} = (x, y, \varphi) \}$ coincides with the motion group of the plane $SE(2)$.

Let $\boldsymbol {v}=(v_1,v_2)$ denote the projections of the velocity of point $R$ onto the moving axes $Rx_1x_2$ relative to the
fixed coordinate
system $Oxy$
and let $\omega$ be the angular velocity of the body. Then
\begin{equation}
\label{eq112}
\dot{x}=v_1\cos\varphi-v_2\sin\varphi, \quad \dot{y}=v_1\sin\varphi+v_2\cos\varphi, \quad \dot{\varphi}=\omega.
\end{equation}
The constraint Eq.~\eqref{eq1} in this case has the form
\begin{equation}
\label{EquationSv}
v_2=0.
\end{equation}

The kinetic energy of the platform can be represented as
$$
T_{\rm s} = \frac{1}{2}m_{\rm s}\big((v_1 - d_2\omega)^2 + (v_2 + d_1\omega)^2 \big) + \frac{1}{2}\big(I_{\rm s} - m_{\rm s}(d_1^2 + d_2^2) \big)\omega^2,
$$
where $m_{\rm s}$ and $I_{\rm s}$ are, respectively, the mass and the moment of inertia of the body relative to the
point of contact $R$, and $\boldsymbol {d}=(d_1,d_2)$ is the radius vector of the center of mass in the moving coordinate system $Rx_1y_1$.

The kinetic energy of the system of material points has the form
$$
T_{\rm p} = \frac{1}{2}\sum_{i=1}^n m_{\rm p}^{(i)}\left(\big( v_1 + \dot{\rho}^{(i)}_1 - \rho^{(i)}_2\omega  \big) ^2  + \big(v_2 + \dot{\rho}^{(i)}_2 +
\rho^{(i)}_1\omega  \big)^2\right) ,
$$
where $m_{\rm p}^{(i)}$ is the mass of the $i$th point.

The kinetic energy of the entire system (platform $+$ material points) can be represented as
$$
\begin{gathered}
T = \frac{1}{2}m\boldsymbol {v}^2 + m\omega\big(c_1(t)v_2  - c_2(t)v_1\big) + \frac{1}{2}I(t)\omega^2  + m \big( v_1\dot{c}_1(t) + v_2\dot{c}_2(t) \big) + k(t)\omega, \\
\end{gathered}
$$
where $m=m_{\rm s} + \sum\limits_{i=1}^n m_{\rm p}^{(i)}$ is the mass of the entire system, $I(t)$ is its moment of inertia,
$\boldsymbol {c}= \big(c_1(t), c_2(t)\big)$ is the position of the center of mass,
and $k(t)$ is the gyrostatic momentum due to the motion of the points.
The last four quantities are given functions of time that are expressed in terms of the system parameters as follows:
\begin{equation}
\label{eq001}
\begin{gathered}
k = \sum_{i=1}^nm_{\rm p}\left( \rho^{(i)}_1 \dot{\rho}^{(i)}_2 - \rho^{(i)}_2 \dot{\rho}^{(i)}_1 \right), \quad
I = I_{\rm s} + \sum_{i=1}^nm_{\rm p}^{(i)}\left( \big(\rho_1^{(i)}\big)^2 + \big(\rho_2^{(i)}\big)^2 \right),\\
c_j = \frac{m_s}{m}d_j + \frac{1}{m}\sum\limits_{i=1}^n m_{\rm p}^{(i)}\rho_j^{(i)}, \quad j=1,2.
\end{gathered}
\end{equation}

Now, for the system at hand, we write the Lagrange equations with undefined multipliers~\cite{BM2}:
\begin{equation}
\label{eq01}
\frac{d}{dt}\left( \frac{\partial T}{\partial \omega} \right) = v_2 \frac{\partial T}{\partial v_1} - v_1 \frac{\partial T}{\partial v_2},\quad
\frac{d}{dt}\left( \frac{\partial T}{\partial v_1} \right) = \omega \frac{\partial T}{\partial v_2}, \quad
\frac{d}{dt}\left( \frac{\partial T}{\partial v_2} \right) = - \omega \frac{\partial T}{\partial v_1} + \lambda,
\end{equation}
where $\lambda$ is the undefined multiplier corresponding to the constraint~\eqref{EquationSv}.
Now we need to restrict these equations and Eq.~\eqref{eq112} to the constraint $v_2=0$ and to eliminate the undetermined
multiplier~$\lambda$. (To eliminate $\lambda$, we have to neglect the last equation in~\eqref{eq01}, since
$\lambda$ does not appear in the other equations.)

We first note that in the case of restriction to the constraint the relation
$$
\left. \frac{\partial T}{\partial v_2}\right|_{v_2=0} = m\big(c_1(t)\omega + \dot{c}_2(t)\big)
$$
is satisfied. Moreover, in the equations of motion it is more convenient to pass from the variables $v_1$ and $\omega$
to new variables, namely, linear momentum $P$ and angular momentum $M$, which are given by
\begin{equation}
\label{eq02}
\begin{gathered}
P = \left. \frac{\partial T}{\partial v_1}\right|_{v_2=0}    \!\! =\! m \big(v_1 - c_2(t)\omega + \dot{c}_1(t) \big), \\
M = \left. \frac{\partial T}{\partial \omega} \right|_{v_2=0}\!\! =\! I(t)\omega - mc_2(t)v_1 + k(t).
\end{gathered}
\end{equation}	

Finally, we obtain equations of motion in the form
\begin{equation}
\label{eq03}
\begin{gathered}
\dot{P} = m\omega\big(c_1(t)\omega + \dot{c}_2(t)\big), \quad
\dot{M} = - m v_1\big(c_1(t)\omega + \dot{c}_2(t)\big), \\
\dot{\varphi} = \omega, \quad \dot{x} = v_1\cos\varphi, \quad \dot{y} = v_1\sin\varphi,
\end{gathered}
\end{equation}	
where for velocities $v_1$ and $\omega$ we find the following expressions from
\eqref{eq02}:
\begin{equation}
\label{eq004}
\begin{gathered}
v_1 = \frac{I(t)P + mc_2(t)M - mI(t)\dot{c}_1(t) - mc_2(t)k(t)}{m\big(I(t) - mc_2^2(t)\big)}, \\ \omega = \frac{c_2(t)P + M  - m c_2(t)\dot{c}_1(t) - k(t)}{I(t) - mc_2^2(t)}.
\end{gathered}
\end{equation}
We note that the denominator in~\eqref{eq004} is a positive definite function.

Thus, regardless of the number of particles, the equations of motion contain four given functions of time:
$c_1(t)$, $c_2(t)$, $I(t)$, $k(t)$.
The resulting system is similar to the well-known Liouville system~\cite{Liouville}, which describes the dynamics of a rigid body
deformable by
a given law.

Equations~\eqref{eq03} are invariant under the motion group of the plane $SE(2)$. As a
result, a closed  (reduced) system of equations decouples which describes the evolution of $P$ and $M$.
It follows from~\eqref{eq03} that the motion of the sleigh in the fixed coordinate system $Oxy$ is defined by
quadratures using the known solutions of the reduced system.

\section{Cases of existence of first integrals and invariant relations}
\label{biz-sec2}
As shown in the previous section, an arbitrary motion of $n$ points reduces to four given functions of time~\eqref{eq001}
in the equations of motion~\eqref{eq03}.
It turns out that for some restrictions to these functions the system~\eqref{eq03} can possess a first integral or
an invariant relation, which allows some conclusions on the dynamics of the sleigh. In this section, we consider these cases
in more detail.

\subsection{A sleigh balanced relative to the knife edge}
\label{BalSl}

Suppose that $c_1(t) \equiv 0$, that is, the center of mass of the system lies on the normal to the plane of the knife edge,
which passes through the point of contact $R$.
In this case, the reduced system reduces to the linear  system
\begin{equation}
\label{eq002}
\begin{gathered}
\dot{P} = \frac{m\dot{c}_2(t)}{I(t) - mc_2^2(t)}\big(c_2(t)P + M - k(t) \big), \\
\dot{M} = - \frac{\dot{c}_2(t)}{I(t) - mc_2^2(t)}\big(I(t)P + mc_2(t)M - mc_2(t) k(t) \big).
\end{gathered}
\end{equation}

These equations possess an additional integral corresponding to the angular momentum of the system
relative to the point of contact:
\begin{equation}
\label{eqL}
L = c_2(t)P + M.
\end{equation}
Let us fix the level set of the integral $L=l$. Then the solution of~\eqref{eq002} can be represented as
$$
\begin{gathered}
P(t) = \widetilde{P}(t) + p_0, \quad M(t) = l - c_2(t)\big(  \widetilde{P}(t) + p_0 \big) \\
\widetilde{P}(t) = m\int\limits_0^t\frac{l - k(\tau)}{I(\tau) - m c_2^2(\tau)}\left( \frac{dc_2}{d\tau} \right) d\tau.
\end{gathered}
$$

Thus, the problem of acceleration or deceleration of the sleigh reduces to investigating the function $\widetilde{P}(t)$.
For given periodic functions $c_2(t)$, $I(t)$ and $k(t)$ this case is examined in more detail in Section~\ref{r1}.

\subsection{Motion along the knife edge}
\label{VL}
Suppose that the material points move on the platform only in the direction of the knife edge:
$$
\rho_2^{(i)} = \const, \quad i=1,\dots,n,
$$
then we find from~\eqref{eq001} that
\begin{equation}
\label{eqVL2}
c_2=\const, \quad k(t) = - mc_2\dot{c}_1(t).
\end{equation}
In this case, it is more convenient to write the equations of motion in the variables $P$ and $L$, where, according to~\eqref{eqL}
and taking~\eqref{eqVL2} into account, we obtain
$$
L=c_2 P+M=\big(I(t)-mc_2^2\big)\omega.
$$
The reduced system can be represented as
$$
\dot{P} = \frac{mc_1(t)}{\big(I(t) - mc_2^2\big)^2} L^2, \quad \dot{L} = -\frac{c_1(t)}{I(t) - mc_2^2} L\big(P - m\dot{c}_1(t) \big).
$$
This system possesses the invariant relation
$$
L=0,
$$
which corresponds to straight-line motion of the sleigh along the knife edge
($\omega=0$).

\begin{remark}
In the general case, the equations of motion~\eqref{eq03} possess the invariant manifold $\omega=0$ if the following relation holds:
\begin{equation}
\label{eqVL}
m\ddot{c}_1(t)c_2(t) + \dot{k}(t)=0.
\end{equation}
Another example for which relation~\eqref{eqVL} holds is considered in Section~\ref{alpha0}.
\end{remark}

\section{Transverse oscillations~--- particular cases}
\label{biz-sec3}
\begin{figure}[!ht]
\centering
		\includegraphics{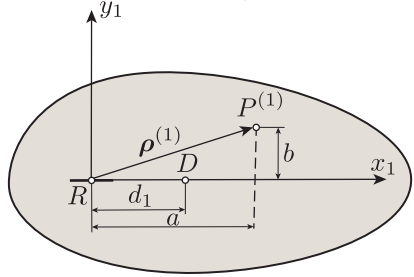}
		\caption{The Chaplygin sleigh.}
		\label{fig06}
\end{figure}

Consider in detail the case where one material point (i.\,e., $n=1$) executes periodic motions on the platform in
a direction transverse to the plane of the knife edge
$$
\boldsymbol {\rho}^{(1)} = \big(a, \ b\sin(\Omega t)\big).
$$
In addition, we shall assume that the center of mass of the platform itself lies on the axis $Rx_1$, that is, $d_2=0$
(see~Fig.~\ref{fig06}).
In this case, from relations~\eqref{eq001} we obtain
\begin{equation}
\label{eqKK}
\begin{gathered}
I(t) = I_{\rm s} + m_{\rm p}^{(1)}\big(a^2 + b^2\sin^2(\Omega t)\big), \quad k(t) = m_{\rm p}^{(1)}ab\Omega\cos(\Omega t), \\
c_1 = \frac{m_{\rm s}}{m}d_1 +  \frac{m_{\rm p}^{(1)}}{m} a, \quad c_2(t) = \frac{m_{\rm p}^{(1)}}{m} b \sin(\Omega t), \quad
m = m_{\rm s} + m_{\rm p}^{(1)}.
\end{gathered}
\end{equation}

Let us define the dimensionless variables $Z_1, Z_2$,  $\tau$, $X$, and
$Y$:
$$
Z_1 = \frac{P}{mb\Omega}, \quad Z_2 = \frac{L}{mb^2\Omega}, \quad \tau = \Omega t,
 \quad X=\frac{x}{b}, \quad Y=\frac{y}{b},
$$
where the angular momentum $L$ is defined by~\eqref{eqL}.
The reduced system of equations in these variables has the form
\begin{equation}
\label{eq04}
\begin{gathered}
\frac{dZ_1}{d\tau} = \frac{\big( Z_2 - \alpha \mu \cos \tau  \big)\big( \delta (Z_2 -  \alpha \mu \cos \tau) + \mu\cos\tau( J + \mu(1 - \mu)\sin^2\tau)\big)}
{\big(J + \mu(1 - \mu)\sin^2\tau\big)^2}, \\
\frac{dZ_2}{d\tau} = - \frac{\delta(Z_2 - \alpha \mu \cos \tau)Z_1}{ J  + \mu(1 - \mu)\sin^2\tau}, \\
\alpha=\frac{a}{b}, \quad \delta = \frac{c_1}{b}, \quad \mu = \frac{m_{\rm p}^{(1)}}{m}, \quad J = \frac{I_{\rm s} + m_{\rm p}^{(1)}a^2}{mb^2}.
\end{gathered}
\end{equation}
The equations of motion for configuration variables are represented as
\begin{equation}
\label{eq0044}
\begin{gathered}
\frac{d\varphi}{d\tau} = \widetilde{\omega}, \quad \frac{dX}{d\tau} = \widetilde{v}_1\cos\varphi, \quad \frac{dY}{d\tau} = \widetilde{v}_1\sin\varphi, \\
\widetilde{\omega} = \frac{ Z_2 - \alpha\mu \cos\tau}{  J + \mu(1 - \mu)\sin^2\tau}, \quad
\widetilde{v}_1 = Z_1+\mu\sin\tau \frac{Z_2-\alpha\cos\tau}{J+\mu(1-\mu)\sin^2\tau}.
\end{gathered}
\end{equation}
We note that, in this case, $0\leqslant \mu<1$ and $J>0$, hence, the denominator in~\eqref{eq04} and~\eqref{eq0044} is
always positive. In addition, the condition that the moment of inertia of the platform relative to the center of mass $D$ be
positive, i.\,e., $I_{\rm s} - m_{\rm s}d_1^2>0$, implies that the inequality restricting the region of physically possible
parameters must be satisfied:
\begin{equation}
\label{eqN}
(1 - \mu)(J - \delta^2) - \mu(\alpha - \delta)^2>0.
\end{equation}

Thus, the problem reduces to investigating the dynamics of the system~\eqref{eq04}, \\~\eqref{eq0044}.
Of particular interest is the question of whether the reduced system~\eqref{eq04} has trajectories unbounded on the plane $(Z_1,
Z_2)$ (i.\,e., trajectories that leave any bounded region on the plane). In this
case, we conclude from~\eqref{eq004} and~\eqref{eqKK} that the velocities of the platform, and hence the kinetic energy, must
increase (in absolute values) indefinitely with time.

We note that the system~\eqref{eq04} has the involution
\begin{equation}
\label{Inv01}
Z_1 \to - Z_1, \ \tau \to -\tau.
\end{equation}
This implies, in particular, that any attractor of the system~\eqref{eqKK} corresponds to a repeller symmetric relative to this
involution.

In what follows, to analyze the position and orientation of the system, it is convenient to use a complex representation
of~\eqref{eq04}:
\begin{equation}
\label{eq15}
\frac{d\varphi}{d\tau}=\frac{Z_2-\alpha \mu \cos\tau}{J+\mu(1-\mu)\sin^2\tau}, \quad
\frac{dz}{d\tau}=(Z_1+i\mu\cos\tau)e^{i\varphi},
\end{equation}
where $z=X+iY+i\mu\sin\tau e^{i\varphi}$.

\subsection{The balanced system $(\delta = 0)$, an additional integral and unbounded acceleration}
\label{r1}
Suppose that the center of mass of the entire system coincides with the point of contact $R$ (that is, $\delta=0$).
In this case, the system~\eqref{eqKK} admits an additional first integral
$$
Z_2=\const.
$$
That is, motion on the plane occurs along the straight lines $Z_2=C_0$ where $C_0=\const$. In this case, the evolution
of the variable $Z_1$ is given by
$$
\frac{d Z_1}{d\tau}=g(\tau), \quad g(\tau)=\frac{\mu(C_0 - \alpha\mu\cos \tau)\cos \tau}{J + \mu(1 - \mu)\sin^2\tau}.
$$

Since $g(\tau)$~is a periodic function, $g(\tau)=g(\tau+2\pi)$, the general solution of this
equation can, as is well known, be represented as
\begin{equation}
\label{eq1714}
Z_1(\tau)=\langle g\rangle \tau+f(\tau),
\end{equation}
where $\langle g\rangle$ is the average over the period of the function $\langle g\rangle$, and $f(\tau)$~is the $2\pi$-periodic
function. In this case,
\begin{equation}
\label{eq1713}
\langle g\rangle=-\frac{\alpha\mu^2}{JB^2}\Big(\sqrt{1+B^2}-1\Big), \quad B^2=\frac{\mu(1-\mu)}{J},
\end{equation}
$$
f(\tau)\!\!=\!\!C_1+\frac{\mu}{JB^2}\!\bigg(\!C_0B\! \arctan(B\sin\tau)+\alpha\mu \sqrt{1+B^2}\Big(\tau-\pi n-\arctan\!\!\big(\sqrt{1\!\!+\!\!B^2}\!\tan\tau\big)\!\Big)\!\bigg),
$$
$$
\tau\in \big(-\pi /2, -\pi/2+\pi n\big), \quad n\in \mathbb{Z}.
$$

Substituting the resulting solutions into~\eqref{eq15}, we obtain an equation governing the evolution of the orientation and
position of the platform in the form
\begin{equation}
\label{eq1712}
\frac{d\varphi}{d\tau}=\frac{C_0-\alpha\mu \cos\tau}{J(1+B^2\sin^2\tau)}, \quad \frac{dz}{d\tau}=\Big(\langle g \rangle\tau+f(\tau)+i\mu\cos\tau\Big)e^{i\varphi}.
\end{equation}

The solution of the first of these equations (as above) can be written as
\begin{equation*}
%\label{eq1734}
\varphi(\tau)=\Omega_0\tau+\Phi(\tau),
\end{equation*}
where $\Omega_0$~is the angular velocity averaged over a period, and $\Phi(\tau)$~is the $2\pi$-periodic function:
$$
\Omega_0=\frac{C_0}{J\sqrt{1+B^2}}, \quad \Phi(\tau)=\frac{C_0}{J\sqrt{1+B^2}}\bigg(\arctan\Big(\sqrt{1+B^2}\tan\tau\Big)-\tau+\pi n\bigg),
$$
$$
\tau\in \bigg(-\frac{\pi}{2},-\frac{\pi}{2}+n\bigg), \quad n\in \mathbb{Z}.
$$

Hence, we conclude that there exist two cases $\langle g\rangle=0$ and  $\langle g\rangle\ne 0$, for which the behavior
of the system
differs qualitatively. Let us consider them in order.

{\bf Case $\langle g\rangle=0$.} Since we assume $\mu\ne 0$\footnote{We recall that, if $\mu=0$, the system is a usual Chaplygin sleigh.},
it follows that $\alpha=0$, that is, the material point oscillates along the axis $Rx_2$ and the center of mass of the
platform $D$
coincides with the point of contact $R$. From relations~\eqref{eq1714},~\eqref{eq1713} and~\eqref{eq1712} we conclude:

\medskip\noindent
{\it if $\delta=0$ and $\alpha=0$, the linear and angular velocities of the platform are bounded
$2\pi$-periodic functions of time $\tau$.}
\medskip

\begin{figure}[!ht]
\centering
		\includegraphics[width=90mm]{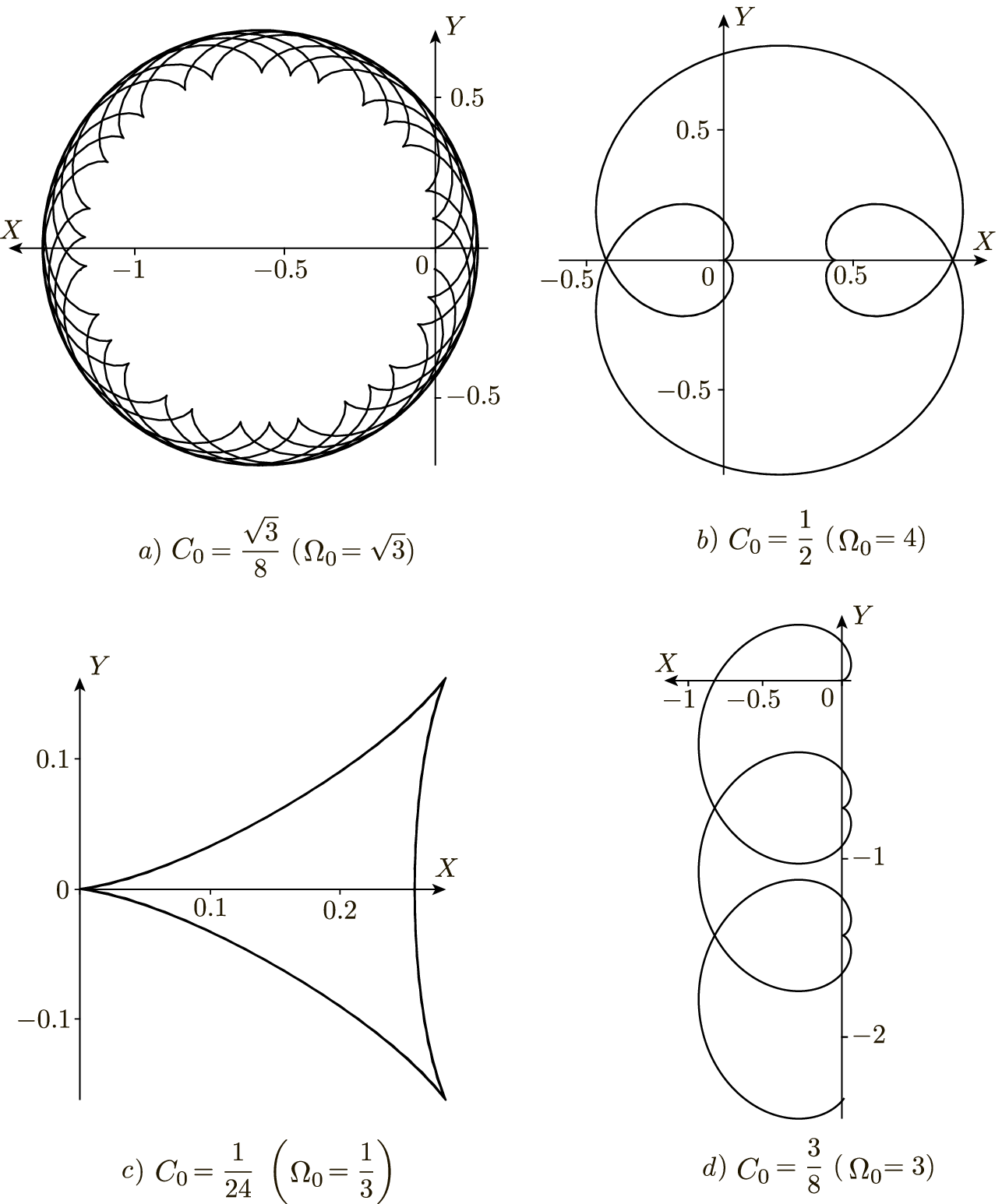}
		\caption{Different trajectories of the point of contact of the sleigh for fixed parameters $\alpha = 0$,  $\delta=0$,
			 $J=\frac{1}{16}$, $\mu=\frac{1}{4}$ and initial conditions $Z_1=0$, $\tau=0$ $\varphi=0$, $X=0$, $Y=0$.}
		\label{fig20}
\end{figure}

In the second equation in~\eqref{eq1712} we expand the periodic functions as a Fourier series and obtain
$$
\frac{dz}{d\tau}=\sum\limits_{m\in \mathbb{Z}}A_m e^{i(\Omega_0+m)\tau},
$$
where $A_m$~are some (complex) numbers.
As is well known (see, e.\,g.,~\cite{AKN}), depending on the value of $\Omega_0$ and the coefficients, three types of
behavior of $z(\tau)$ and hence of the point of contact $R$ of the platform are possible.
\begin{itemize}
\item[{$1^\circ$}\!\!.] Let $\Omega_0$~be irrational, then the trajectory of point $R$ is a nonclosed curve consisting of
equal segments traced out over period $2\pi$ (lobes~\cite{BM, BKM}), which for each subsequent period
rotate relative to some center by an angle that is incommensurate with angle $2\pi$, see Fig.~\ref{fig20}a.
\item[{$2^\circ$}\!\!.] If $\Omega_0=\frac{p}{q}$, $p,\, q\in \mathbb{Z}$ (but is not an integer), then the trajectory
of point $R$ turns out to be $2\pi q$-periodic, see Fig.~\ref{fig20}c.
\item[{$3^\circ$}\!\!.] Let $\Omega_0=2m_*+1$, $m_*\in \mathbb{Z}$ and $A_{m_*}\ne 0$, then
$$
z(\tau)=A_{m_*}\tau+\tilde{z}(\tau)=O(\tau),
$$
where $\tilde{z}(\tau)$~is the $2\pi$-periodic function. In this case, the trajectory of point $R$ is unbounded, and
the displacement for a period is some
fixed value, see Fig.~\ref{fig20}d. If $\Omega_0=2m_*$, then in $A_{m_*} = 0$  (since in the second equation
of~\eqref{eq1712} $f(\tau)$ is an odd function of time).
As a result, the trajectory of point $R$ turns out to be periodic, see Fig.~\ref{fig20}b.
\end{itemize}

{\bf Case $\langle g\rangle\ne 0$.} We first note that in this case, according to~\eqref{eq1714}, an unbounded
acceleration of the platform is observed.

\begin{propos}
For $\delta=0$ and $\alpha\ne 0$, the linear velocity of the platform increases indefinitely $($linearly in time$)$, whereas the
angular velocity remains
bounded.
\end{propos}

This also implies that the kinetic energy of the system also increases indefinitely (quadratically in time).

To analyze the behavior of the point of contact $R$, we also expand all periodic functions on the right-hand side of the second
equation in~\eqref{eq1712} and obtain
\begin{equation}
\label{eq18}
\frac{dz}{d\tau}=\sum\limits_{m\in \mathbb{Z}}(A_m+B_m\tau)e^{i(\Omega_0+m)\tau}.
\end{equation}
As in the previous case, there are two situations in which qualitative differences in the behavior
of the platform can be observed.
\begin{itemize}
\item[{$1^\circ$\!\!.}] $\Omega_0\in \mathbb{Z}$, then integrating~\eqref{eq18}, we obtain
$$
z(\tau)=\sum\limits_{m\in \mathbb{Z}}\left(\frac{A_m+B_m \tau}{i(\Omega_0+m)}+\frac{B_m}{(\Omega_0+m)^2}\right)e^{i(\Omega_0+m)\tau}=O(\tau).
$$
\item[{$2^\circ$\!\!.}] For some $m_*\in \mathbb{Z}$, $\Omega_0=m_*$ and $B_{m_*}\ne 0$, then
$$
z(\tau)=\frac{1}{2}B_{m_*}\tau^2+O(\tau).
$$
\end{itemize}

\begin{figure}[!ht]
\centering
		\includegraphics[totalheight=4.5cm]{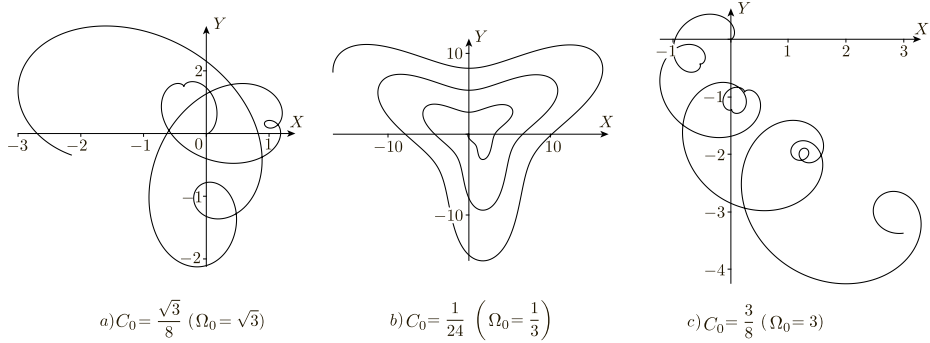}
		\caption{Different trajectories of the point of contact of the sleigh for fixed parameters $\alpha = \frac{1}{3}$,
$\delta=0$,
			$J=\frac{1}{16}$, $\mu=\frac{1}{4}$ and initial conditions $Z_1=0$, $\tau=0$ $\varphi=0$, $X=0$, $Y=0$.}
		\label{fig07}
\end{figure}

\begin{figure}[!ht]
\centering
		\includegraphics[totalheight=4.5cm]{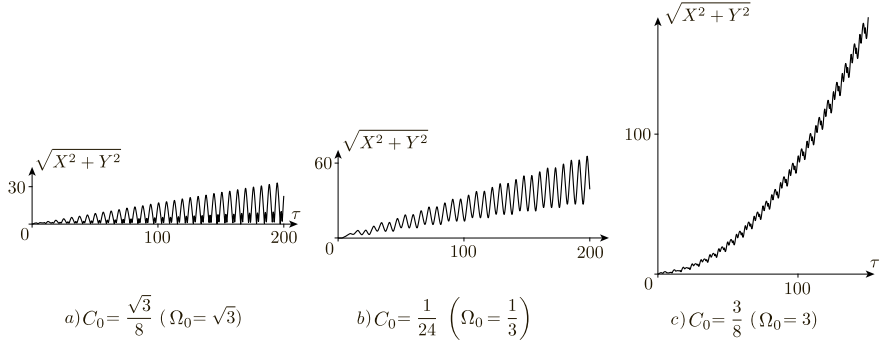}
		\caption{The dependence of $\sqrt{X^2 + Y^2}$ on $\tau$ for fixed parameters $\alpha = \frac{1}{3}$,  $\delta=0$,
			$J=\frac{1}{16}$, $\mu=\frac{1}{4}$ and initial conditions $Z_1=0$, $\tau=0$ $\varphi=0$, $X=0$, $Y=0$.}
		\label{fig25}
\end{figure}

In both cases, the trajectory of the platform is unbounded, but the velocity of motion from the initial point is different (see Figs.~\ref{fig07},~\ref{fig25}).

\subsection{The case of existence of a singular invariant measure}
\label{alpha0}
Set
$$
\alpha=0, \quad \delta\ne 0,
$$
that is, the oscillating mass moves along the axis $Ry_1$, but the center of mass $D$ does not coincide with the point of contact
$R$. In this case, the reduced system can be represented as
\begin{equation}
\label{ZZ1}
\frac{dZ_1}{d\tau}= \frac{Z_2\Big( \delta Z_2 + \mu\cos\tau\big( J + \mu(1 - \mu)\sin^2\tau \big)  \Big)}{\big( J + \mu(1 - \mu)\sin^2\tau\big)^2}, \
\frac{dZ_2}{d\tau}= - \frac{\delta Z_2Z_1}{J + \mu(1 - \mu)\sin^2\tau}.
\end{equation}
We first note that in Eqs.~\eqref{ZZ1} by making the change of variables
$$
Z_1\to \frac{1}{\delta}Z_1, \quad Z_2\to \frac{1}{\delta}Z_2
$$
we can eliminate the parameter $\delta$.
However, by analogy with the other section, we keep $\delta$ in this system, but assume
$$
\delta>0.
$$

The system~\eqref{ZZ1} possesses a {\it singular} invariant measure with density
$$
\rho=\frac{1}{Z_2}.
$$

The density of the invariant measure $\rho$ has a singularity on the submanifold
$$
\Sigma_s=\{ (Z_1, Z_2), \, Z_2 = 0 \},
$$
which coincides with the abscissa axis.
As is well known, this submanifold is invariant (for details, see~\cite{BizBM,Koz2016}), and, according to~\eqref{ZZ1},
is filled with fixed points in this case.
This implies, in particular, that any trajectory of the system~\eqref{ZZ1} cannot cross the straight line
$\Sigma_s$. Therefore, we restrict our attention to the trajectories in the upper half-plane
$$
Z_2>0.
$$
(For $Z_2<0$ the trajectories can be obtained by making the change of variables $Z_2\to -Z_2$ and by rescaling time by $\tau\to \tau+\pi$.)

Let us calculate the divergence of the vector field~\eqref{ZZ1}:
$$
D = - \frac{\delta Z_1}{J + \mu(1 - \mu)\sin^2\tau}.
$$
We see that when $Z_1>0$, the flow~\eqref{ZZ1} compresses the phase volume, and when $Z_1<0$, the volume is expanded. We now show
rigorously that the equilibrium points lying on $\Sigma_s$ are asymptotically stable for $Z_1>0$ and \\ asymptotically unstable
for $Z_1<0$. To do so, we make use of the Lyapunov method for constructing the functions $F(Z_1, Z_2)$ whose
derivatives along the trajectories of the system retain their sign.
Thus, their level surfaces bound  possible trajectories of the system in a natural way.

To find these functions, we consider at each point of the plane $(Z_1, Z_2)$ the tangent of the angle
between the axis $OZ_2$ and the vector field~\eqref{ZZ1}
$$
A(Z_1, Z_2)=\left( \frac{dZ_2}{d\tau}\right)^{-1}\frac{dZ_1}{d\tau}=-\frac{1}{Z_1}\left( \frac{Z_2}{J + \mu(1-\mu)\sin^2\tau} + \frac{\mu}{\delta}\cos\tau\right) .
$$
We see that the vector field is bounded from above and below as follows:
$$
\begin{gathered}
-\frac{1}{Z_1}\left( \frac{Z_2}{J} + k_+\frac{\mu}{\delta}\right)
\leqslant A(Z_1, Z_2) \leqslant  - \frac{1}{Z_1}\left( \frac{Z_2}{J + \mu(1 - \mu)} - k_-\frac{\mu}{\delta}\right), \ \mbox{for} \ Z_1>0, \\
 - \frac{1}{Z_1}\left( \frac{Z_2}{J + \mu(1 - \mu)} - k_-\frac{\mu}{\delta}\right)  \leqslant A(Z_1, Z_2) \leqslant -\frac{1}{Z_1}
\left( \frac{Z_2}{J} + k_+\frac{\mu}{\delta}\right), \ \mbox{for} \ Z_1<0,
\end{gathered}
$$
where $k_+$ and $k_-$ are some constants satisfying the inequalities $k_+ > 1$ and $k_- > 1$.
This implies that at each point $(Z_1,Z_2)$
at all instants of time $\tau$ the vector field of the system~\eqref{ZZ1} is contained between the pairs of straight lines given by
distributions of the form
$$
Z_1dZ_1 + \left( \frac{Z_2}{J} + k_+\frac{\mu}{\delta} \right)dZ_2=0, \quad Z_1dZ_1 + \left( \frac{Z_2}{J + \mu(1-\mu)} - k_-\frac{\mu}{\delta} \right) dZ_2=0.
$$
Their integrals yield the required functions
\begin{equation}
\label{eq2312}
\begin{gathered}
F_+ = \frac{1}{2}Z_1^2 + \frac{1}{2J}\left( Z_2 + k_+\frac{\mu J}{\delta} \right)^2, \\
F_-=\frac{1}{2}Z_1^2 + \frac{1}{2\big(J + \mu(1-\mu)\big)}\left(Z_2 - k_-\frac{\mu\big(J+\mu(1-\mu)\big)}{\delta} \right)^2.
\end{gathered}
\end{equation}
Differentiating these functions along the system~\eqref{ZZ1}, we find
\begin{equation*}
\begin{gathered}
\dot{F}_+=-\mu Z_1Z_2\frac{J\big(k_+ - \cos\tau\big)\big(J+\mu(1-\mu)\sin^2\tau\big) +\delta Z_2(1 - \mu)\sin^2\tau}
{J\big(J+\mu(1-\mu)\sin^2\tau\big)^2} \\
\dot{F}_-=\mu Z_1Z_2\frac{\mu\big(J + \mu(1 - \mu)\big)\big(k_- + \cos\tau\big)\big(J+\mu(1-\mu)\sin^2\tau\big) +\delta Z_2(1 - \mu)\cos^2\tau}
{\big(J + \mu(1 - \mu) \big)\big(J+\mu(1-\mu)\sin^2\tau\big)^2}.
\end{gathered}
\end{equation*}

\begin{figure}[!ht]
\centering
		\includegraphics[totalheight=8cm]{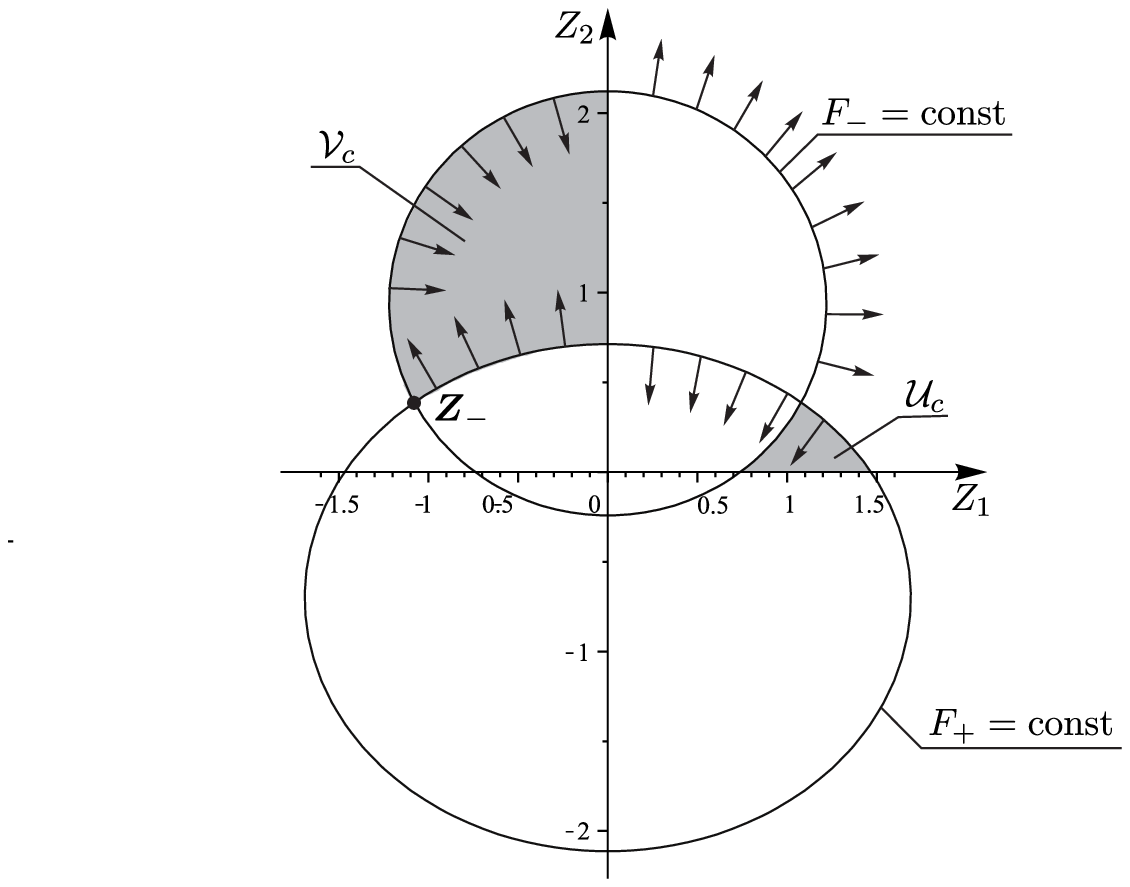}
		\caption{Level lines of the functions~\eqref{eq2312}. The arrows show the direction of motion of the level lines of
these functions by virtue of the system~\eqref{ZZ1}.}
		\label{fig27}
\end{figure}
We see that the following inequalities are satisfied:
\begin{equation}
\label{eq2412}
\begin{gathered}
\dot{F}_- >0, \quad \dot{F}_+ <0, \quad \text{for} \quad Z_1>0, \\
\dot{F}_- <0, \quad \dot{F}_+ >0, \quad \text{for} \quad Z_1<0.
\end{gathered}
\end{equation}

This implies, in particular, that any nonempty set formed by the intersection lines of the functions $F_+$, $F_-$, $Z_2$
with $Z_1>0$ (see Fig.~\ref{fig27})
$$
\mathcal{U}_c=\{(Z_1, Z_2)\,|\, F_+\leqslant c_+, \,\, F_-\geqslant c_-, \,\, Z_2\geqslant 0\},
$$
where $c_+$ and $c_-$~are some constants, decreases until it shrinks to a point. On the other hand, when
$Z_1<0$, the trajectory leaves the neighborhood of any point $\boldsymbol Z_-$ (see Fig.~\ref{fig27}). Thus, the following
statement holds.
\begin{propos}
Fixed points lying on the straight line $\Sigma_s$ are unstable for $Z_1<0$ and asymptotically stable for
$Z_1>0$.
\end{propos}
\begin{remark}
A rigorous proof requires that we consider trajectories for $Z_2<0$, but, as noted above, they are obtained from
trajectories for $Z_2>0$ by the transformations $Z_2\to -Z_2$ and $\tau\to \tau+\pi$.
Moreover, due to the involution~\eqref{Inv01}, for the proof it sufficed to show that when
$Z_1>0$, the points on the straight line $\Sigma_s$ are asymptotically stable.
\end{remark}

We note that the equilibrium points $\Sigma_s$ correspond to straight-line motion of the sleigh, since the
equation $\widetilde{\omega} = 0$ holds.

In addition, it follows from conditions~\eqref{eq2412} that for the initial data from the set of points of the form
(see Fig.~\ref{fig27})
$$
\mathcal{V}_c=\{(Z_1, Z_2)\,|\, Z_1\leqslant 0, \,\, F_+\geqslant c_+, \,\, F_-\leqslant c_-\},
$$
where $c_+$ and $c_-$~are some constants, any trajectory of the system~\eqref{ZZ1} remains bounded for $Z_1<0$.

In view of the fact that the coordinate $Z_2(\tau)$ decreases monotonically for $Z_1>0$, we conclude that

\medskip\noindent
{\it  all trajectories of the system~\eqref{ZZ1} are also bounded as
$t\to +\infty\,\,(t\to -\infty)$ and tend to fixed points
on the straight line $\Sigma_s$ for $Z_1>0$ $($for $Z_1<0$$)$}.
\medskip

\begin{figure}[!ht]
\centering
		\includegraphics[totalheight=3cm]{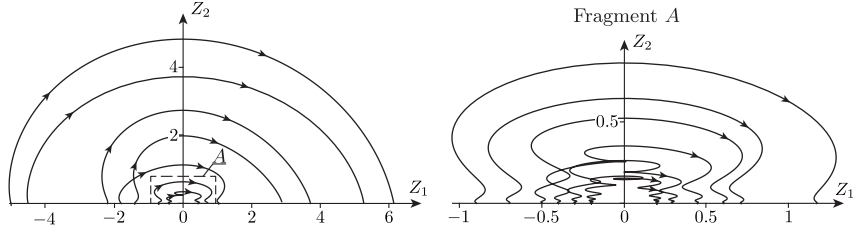}
		\caption{ Trajectories of the system~\eqref{ZZ1} for fixed parameters $\delta=0.3$, $\mu=0.6$, $J=0.7$.}
		\label{Scatt02}
\end{figure}

\begin{figure}[!ht]
\centering
		\includegraphics[totalheight=4.5cm]{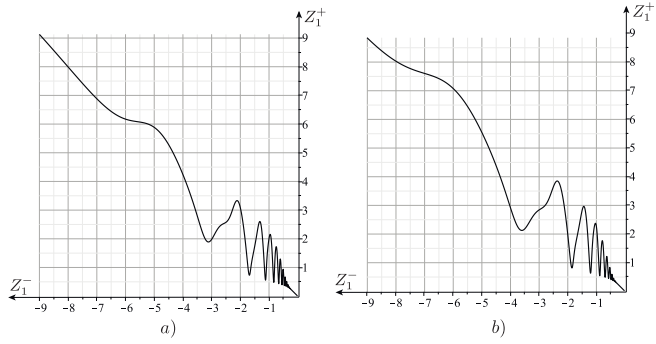}
		\caption{ The dependence $Z_1^+ (Z_1^-)$ for fixed parameters $\delta=0.3$, $\mu=0.6$, $J=0.7$, $\varepsilon = 10^{-7}$
and different $a)$~$\tau_0=0$, $b)$ $\tau_0=\frac{\pi}{5}$.}
		\label{fig16}
\end{figure}

\begin{figure}[!ht]
\centering
		\includegraphics[totalheight=3.5cm]{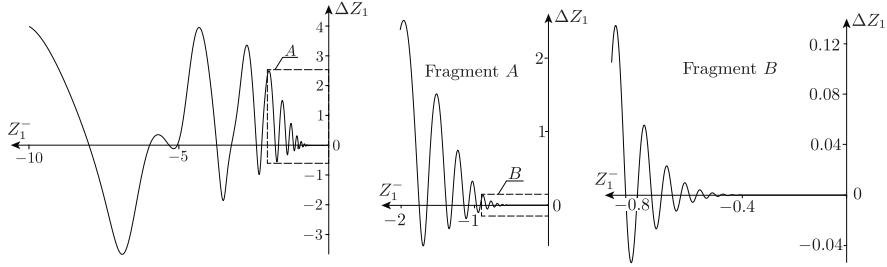}
		\caption{ The dependence $\Delta Z_1(Z_1^{-})$ for fixed parameters $\delta=0.3$, $\mu=0.6$, $J=0.7$, $\varepsilon = 10^{-7}$, $\tau_0=0$.}
		\label{Scatt}
\end{figure}

Thus, in the system~\eqref{ZZ1} any trajectory from a neighborhood of the unstable equilibrium point $\Sigma_s$
with fixed $Z_1 = Z_1^-<0$ as $t\to +\infty$ asymptotically tends to another equilibrium point
corresponding to $Z_1 =Z_1^+>0$ (see Fig.~\ref{Scatt02}).
Therefore, we consider the following family of one-dimensional maps:
$$
\Pi_{\varepsilon, \tau_0}: Z_1^-\to Z_1^+,
$$
where $Z_1(\tau_0)=Z_1^{-}$ and $Z_2(\tau_0)=\varepsilon$ are the initial conditions for the trajectory of the
system~\eqref{ZZ1} and $Z_1^+$ is the value of $Z_1(\tau)$ for this trajectory as $\tau\to\infty$.

It can be seen in Fig.~\ref{fig16} that this map depends considerably on $\tau_0$, and for large absolute values $Z_1^{\pm}$
the map $\Pi_{\varepsilon, \tau_0}$ is similar to a map of the form
\begin{equation}
\label{eq2413}
Z_1^+=-Z_1^-.
\end{equation}

Figure~\ref{Scatt} shows the deviation $\Delta Z(Z_1^-)=Z_1^+ + Z_1^-$ of the scattering map $\Pi_{\varepsilon, \tau_0}$
from the symmetry~\eqref{eq2413}. It is clearly seen that the frequency of oscillations of the function $\Delta Z$
increases indefinitely as $Z_1^-\to 0$. This suggests that this function is not analytic at the point $Z_1^-=0$
(by analogy with functions of the form $x\sin\frac{1}{x}$).

We note that the problem of correctly defining the scattering map for this system remains open. Apparently,
one also needs to consider the phase, and the map does not reduce to a one-dimensional one.
There is an extensive literature devoted to investigating various scattering maps (see, e.\,g.,~\cite{JungScatt, Eckhardt, Aref}).

\section{Transverse oscillations~--- the general case:\\ acceleration and chaotic dynamics}
\label{biz-sec4}
{\bf 1.} We now turn to considering the general case. In this section, we assume without loss of
generality that
$$
\alpha\ne 0, \quad \delta>0.
$$
We recall that the last condition can be satisfied by simultaneously changing the signs of the variables $Z_1$ and $Z_2$ and by
rescaling time by  $\tau\to \tau+\pi$. In addition, the fact that the system has involution~\eqref{Inv01} implies
that, as $\tau\to +\infty$ and $\tau\to -\infty$, the behavior of the trajectories is identical, up to sign, to $Z_1$.

In the general case, the system~\eqref{eq04} has no additional tensor invariants (first integral, invariant measure).
Therefore, it is natural to start its analysis with numerical experiments. Since the dependence on time is periodic is this case,
the system~\eqref{eq04} generates the Poincar\'{e} map of the plane after each period.
In the figures that show the Poincar\'{e} map of this system, periodic solutions correspond to fixed points, and invariant tori
correspond to invariant curves.

\begin{figure}[!ht]
\centering
		\includegraphics[totalheight=2.8cm]{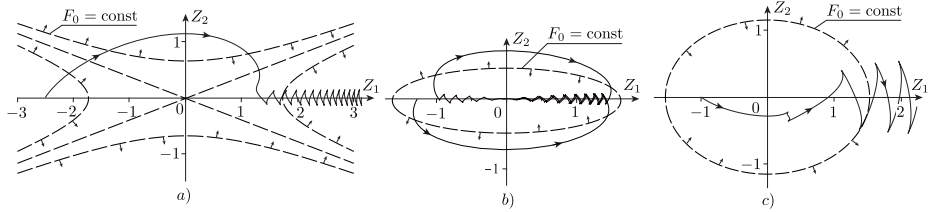}
		\caption{Level lines of $F_0$ and the projection of trajectories with
			fixed parameters $\delta = 0.3$,  $\mu=0.3$,  $J=0.25$  and different $a)\alpha=-0.5$,  $b)\alpha=0.3$, $c)\alpha=2$ onto the plane $Z_1, Z_2$.
			The arrows show the direction of motion of the level lines of $F_0$ by virtue of the system~\eqref{eq04}
and the direction of the trajectories.}
		\label{fig42}
\end{figure}

Depending on the parameters, the system trajectories exhibit the following qualitatively different behaviors.
\begin{itemize}
\item[{1.}] Acceleration~--- all trajectories of the system are noncompact. In this case, $Z_1\to +\infty$ as $\tau\to +\infty$ (see~Fig.~\ref{fig42}c).
\item[{2.}] Stability and multistability~--- all trajectories tend to one or several periodic solutions as \\ $\tau\to +\infty$
(see~Fig.~\ref{chart}d).
\item[{3.}] Chaotic and quasi-periodic oscillations~--- the system has a strange attractor (see~Fig.~\ref{fig22}a), which can
coexist with invariant tori (see~Fig.~\ref{fig50}a).
\end{itemize}
We discuss the question as to for what parameters these situations can arise.

{\bf 2.} We first note that the following estimates hold for the derivatives in~\eqref{eq04}:
\begin{equation}
\label{eq32}
\begin{gathered}
\frac{dZ_1}{d\tau}>0, \quad \text{for} \quad |Z_2|>Z^*_2, \quad Z^*_2=\max \, \left(\mu|\alpha|, \mu\bigg|\alpha-\frac{J+\mu(1 - \mu)}{\delta}\bigg|\right), \\
\frac{d}{d\tau}|Z_2|<0, \quad \text{for} \quad Z_1>0, \quad |Z_2|>\mu|\alpha|, \\
\frac{d}{d\tau}|Z_2|>0, \quad \text{for} \quad Z_1<0, \quad |Z_2|>\mu|\alpha|.
\end{gathered}
\end{equation}

Thus, we see  that at any parameter values and arbitrary instants of time all trajectories of the system {\it outside the strip}
$|Z_2|<Z^*_2$ are directed to the right (that is, $Z_1$ increases), and when $Z_1>0$, the trajectories approach this strip, while when $Z_1<0$, they move away from it.

To analyze possible behavior {\it inside the strip}
$|Z_2|<Z^*_2$, we use, as in the previous section, the method of Lyapunov functions.
In this case, we consider the homogeneous quadratic function
\begin{equation}
\label{eq30}
F_0=Z_1^2 + \frac{Z_2^2}{\alpha\delta}.
\end{equation}
Differentiating it with respect to time using~\eqref{eq04}, we obtain
\begin{equation}
\label{eq31}
\frac{dF_0}{d\tau}=-\frac{2Z_1\big( Z_2  - \alpha\mu\cos\tau \big)^2\big( J - \alpha\delta + \mu(1-\mu)\sin^2\tau \big)}{\alpha\big( J + \mu(1 - \mu)\sin^2\tau \big)^2}.
\end{equation}

\goodbreak

It follows from~\eqref{eq30} and~\eqref{eq31} that, depending on the value of $\alpha$, there are four qualitatively
different situations (see Fig.~\ref{fig42}):
\begin{itemize}
\item[{$\bullet$}] $\alpha<0$, the level lines of $F_0$~are hyperbolas, along the system trajectories the function $F_0$
increases strictly for $Z_1>0$, and decreases strictly for $Z_1<0$
(see Fig.~\ref{fig42}a);
\item[${\bullet}$] $0<\alpha < \frac{J}{\delta}$, the level lines of $F_0$~are ellipses, along the
trajectories the function $F_0$ decreases strictly for $Z_1>0$ and increases strictly for $Z_1<0$, see Fig.~\ref{fig42}b);
\item[{$\bullet$}] $\frac{J}{\delta}<\alpha<\frac{J+\mu(1-\mu)}{\delta}$, the sign of the derivative of $F_0$ is
not defined;
\item[{$\bullet$}] $\frac{J+\mu(1-\mu)}{\delta}<\alpha$, the level lines of $F_0$~are ellipses,
along the trajectories the function $F_0$ increases strictly for $Z_1>0$ and decreases strictly for $Z_1<0$, see Fig.~\ref{fig42}c).
\end{itemize}

\begin{figure}[!ht]
\centering
		\includegraphics[totalheight=4cm]{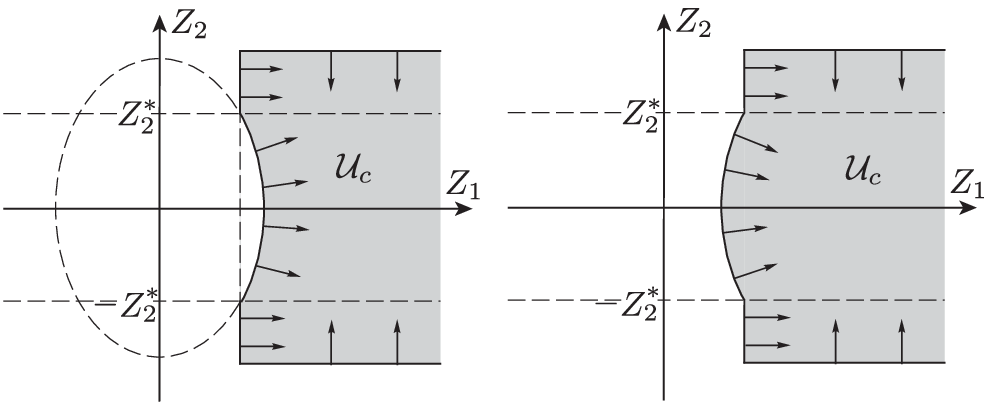}
		\caption{ }
		\label{fig41}
\end{figure}

Assume that $\alpha<0$ or $\frac{J+\mu(1-\mu)}{\delta}<\alpha$. Then inside the strip $|Z_2|<Z^*_2$ the flow of the
system~\eqref{eq04} is directed to the right (towards increase of $Z_1$) at almost all instants of time.
Consider on the plane $(Z_1, Z_2)$ a set
given by the relations (see Fig.~\ref{fig41}):
\begin{equation*}
\begin{gathered}
\mathcal{U}_c=\left\{(Z_1, Z_2)\,\big|\, Z_1>c_1, \, |Z_2|<Z^*_2+c_2,\, F_0 >c^2_1 + \frac{(Z^*_2)^2}{\alpha \delta}\right\}, \quad \text{for} \quad \alpha >0, \\
\mathcal{U}_c=\left\{(Z_1, Z_2)\,\big|\, Z_1>c_1, \, |Z_2|<Z^*_2+c_2,\, F_0 >c^2_1-\frac{(Z^*_2)^2}{|\alpha|\delta}\right\}, \quad \text{for} \quad \alpha <0.
\end{gathered}
\end{equation*}
According to~\eqref{eq31} and~\eqref{eq32}, the trajectories starting at $\tau=\tau_0$ in $\mathcal{U}_c$ remain in this region
for all $\tau>\tau_0$.\goodbreak

Hence, we conclude:

\smallskip
\noindent{\it acceleration in the system~\eqref{eq04} is possible if its parameters satisfy one of the following
conditions:
\begin{equation}
\label{eq33}
\alpha< 0, \quad \text{or} \quad \frac{J+\mu(1-\mu)}{\delta}<\alpha.
\end{equation}
In this case, the linear velocity of the platform increases indefinitely, and the angular velocity remains bounded.}
\smallskip

As computer experiments show, relation~\eqref{eq33} defines sufficient conditions for acceleration
in the system~\eqref{eq04}. However, a rigorous proof of this fact requires more detailed estimates and remains an open problem.
\begin{remark}
To prove that the trajectories outside the strip $|Z_2|<Z^*_2$ are bounded for $Z_1<0$, we need to use a function of the form
$$
F_{\pm}=Z^2_1+\frac{1}{J+\mu(1-\mu)}\left(Z_2\pm k\mu \frac{|\alpha|\delta+J+\mu(1-\mu)}{\delta}\right)^2, \quad k>1.
$$
\end{remark}

{\bf 3.} If we set $0<\alpha<\frac{J}{\delta}$, then, according to~\eqref{eq31} (see Fig.~\ref{fig42}$b$), the trajectories
inside the strip $|Z_2|<Z^*_2$ are directed to the left (that is, towards decrease of $Z_1$). When
$\frac{J}{\delta}<\alpha <\frac{J+\mu(1-\mu)}{\delta}$, there is no definite direction of motion inside the strip.
It is in these cases that the above-mentioned regimes $2$ and $3$ arise.
We illustrate this by numerical analysis of the system.

To carry out numerical analysis of the qualitative behavior of the system for  $0<\alpha<\frac{J+\mu(1-\mu)}{\delta}$, we
specify two parameters
\begin{equation*}
\delta = 0.3, \quad J=0.25.
\end{equation*}
On the parameter plane $(\mu, \alpha)$, we plot a chart of the largest Lyapunov exponent $\lambda_1$ for one trajectory of
the system~\eqref{eq04} (see~Fig.~\ref{chart}e). We note that the region with physical parameter values~\eqref{eqN}
lies on the left of the curve $(1 - \mu)(J - \delta^2) - \mu(\alpha - \delta)^2=0$.

\begin{figure}[!ht]
\vspace{-2mm}
\centering		
\includegraphics[width=\textwidth]{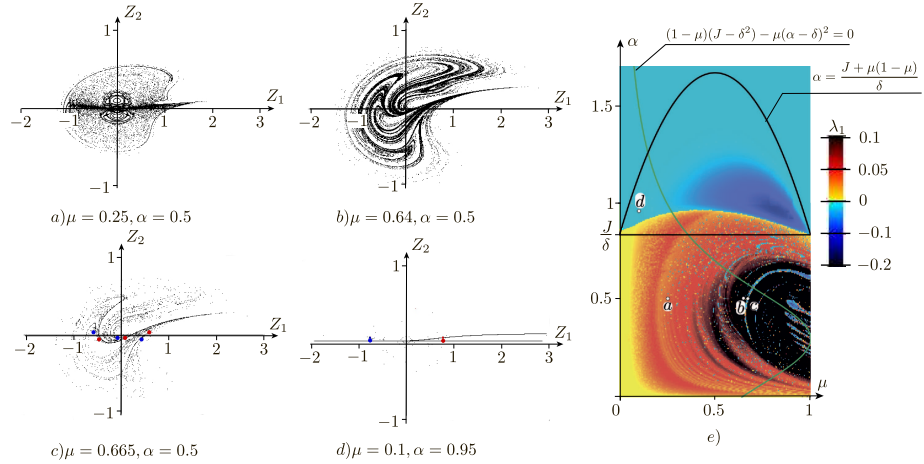}
		\caption{Chart of the largest Lyapunov exponent $\lambda_1$ for the point with  initial conditions $Z_1=5$, $Z_2=0$, $\tau=0$
and iterations of a point map.}
		\label{chart}
\vspace{-2mm}
\end{figure}

As is well known, the largest Lyapunov exponent $\lambda_1$ characterizes the degree of exponential divergence of close trajectories.
If the trajectories are bounded, then the case $\lambda_1>0$ corresponds to chaotic motion, and the case
$\lambda_1\leqslant 0$ corresponds to regular motion.
Figures~\ref{chart}a--\ref{chart}d presents the results of iteration of a point map
after each period $\tau=2\pi$ on the plane $(Z_1,Z_2)$ for different points in Fig.~\ref{chart}e.

As is seen, for $\alpha<\frac{J}{\delta}$ there are both chaotic and regular regimes of motion.
For example, in Fig.~\ref{chart}a one can see, in addition to chaotic trajectories, invariant curves corresponding to
quasi-periodic motion.
The projection of two trajectories of the system~\eqref{eq04} onto the plane $(Z_1,Z_2)$ and the motion of the point of contact
in this case are shown in Fig.~\ref{fig50}.
\begin{figure}[!ht]
\centering
		\includegraphics[totalheight=4cm]{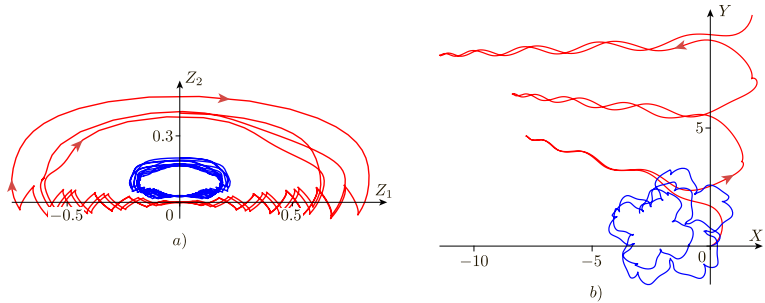}
		\caption{Trajectories of the system \eqref{eq04} and motion of the point of contact for parameters
			corresponding to Fig.\,\ref{chart}a ($\alpha=0.5$, $\delta=0.3$, $\mu=0.25$, $J=0.25$). One of the trajectories
			($\tau=0, Z_1=0, Z_2=0.4$) corresponds to the chaotic motion regime, and the other trajectory
($\tau=0, Z_1=0, Z_2=0.2$) corresponds to quasi-periodic motion.
			 For both trajectories we have chosen
			$Z_1=0$, $\tau=0$, $\varphi=0$, $X=0$, $Y=0$.}
		\label{fig50}
\end{figure}

Figure~\ref{chart}b presents the results of iteration of a chaotic trajectory on a strange attractor
for which the Lyapunov exponents have the form
$$
\lambda_1\approx0.11, \quad \lambda_2\approx0, \quad \lambda_3\approx-0.26.
$$
A typical trajectory of the point of contact of the sleigh on a strange attractor is presented in Fig.~\ref{fig22}.
As shown in~\cite{Kuznetsov},  chaotic dynamics on the attractor of a reduced system
leads in this case to isotropic random motion of the sleigh of diffusion type in the fixed reference system
(with loss of the memory of the initial orientation for large time scales).
A qualitative estimate of this can be given by a graphic representation of an ensemble of segments of the same trajectory
of the sleigh where these segments are displaced on the plane so that the initial points
coincide.
In terms of quantitative statistics, the distribution of distances from the beginning to the end of each segment of
the trajectory, which is achieved at a fixed number of periods of oscillations of the internal mass $N$ (number of
iterations of the Poincar\'{e} map), must asymptotically tend to the Rayleigh distribution, and the azimuth angles must tend to
uniform distribution in an interval from $0$ to $2\pi$~\cite{Biz-A1,Biz-A2,Biz-A3}.
In~\cite{Kuznetsov}, this is illustrated by cumulative distributions for distances and angles which for $N>102$
are in good agreement with theoretical distributions for isotropic
random walks. Using the well-known relation from the theory of two-dimensional random walks, one can estimate the coefficient
of diffusion as a ratio of the half-sum of dispersions for displacements of the sleigh along the axes of the coordinates $Ox$ and $Oy$ to the value of the time
interval.

\begin{figure}[!ht]
\centering
		\includegraphics[totalheight=5cm]{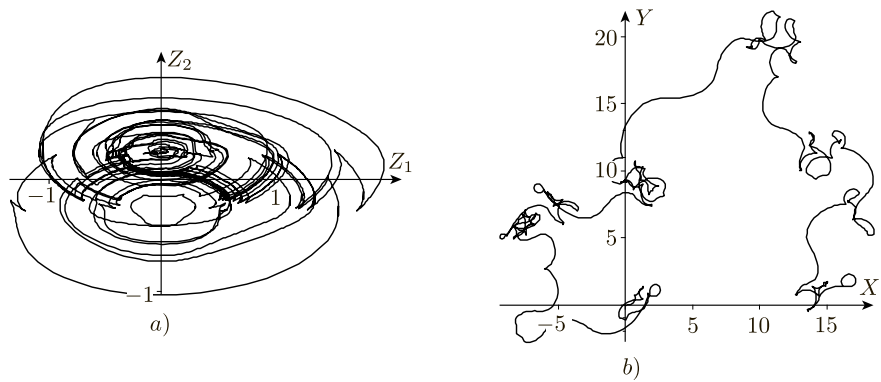}
		\caption{Trajectories of the system \eqref{eq04} and of the point of contact for parameters
			corresponding to Fig.\,\ref{chart}b ($\alpha=0.5$, $\delta=0.3$, $\mu=0.64$, $J=0.25$).
			 The trajectories are plotted for the initial conditions
			 $\tau=0$, $Z_1=0$, $Z_2=0.3$, $\varphi=0$, $X=0$, $Y=0$ and correspond to a strange attractor.}
		\label{fig22}
\end{figure}

\begin{figure}[!ht]
\centering
\includegraphics{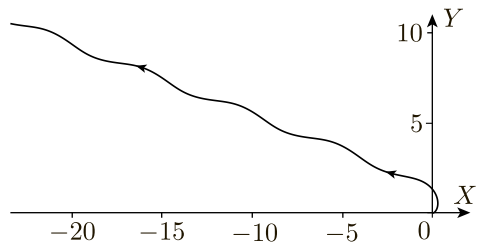}
\caption{Motion of the point of contact at parameter values corresponding to Fig.~\ref{chart}d ($\alpha=0.95$, $\delta=0.3$, $\mu=0.1$, $J=0.25$).
The trajectory is plotted for the initial conditions
$\tau=0$, $Z_1=0$, $Z_2=1$, $\varphi=0$, $X=0$, $Y=0$  and corresponds to a trajectory asymptotically tending to the
limit cycle on the map.}
\label{fig55}
\end{figure}

As is seen from Fig.~\ref{chart}c, in the region $\lambda_1>0$ there are areas for which $\lambda_1<0$. An iteration of the map
for such an area is shown in Fig.~\ref{chart}c, in this case this map has asymptotically stable and unstable high-period fixed points.

After the curve $\lambda_1=0$ is crossed in the region $ \frac{J}{\delta}<\alpha<\frac{J+\mu(1-\mu)}{\delta}$, stable and
unstable  degenerate fixed points\footnote{In Fig.~\ref{chart}d these points are shown in red and blue, respectively.}
arise on the map (see~Fig.~\ref{chart}d).
All trajectories (except for fixed points)
asymptotically tend to a stable point. A typical motion of the point of contact in this case is presented in Fig.~\ref{fig55}.
Further, as $\alpha$ increases, degenerate fixed points move away from the origin (go to infinity)
and, as a result, when $\alpha> \frac{J+\mu(1-\mu)}{\delta}$, acceleration is observed.

The authors express their gratitude to V.\,V.\,Kozlov, S.\,P.\,Kuznetsov and D.\,V.\,Treschev for fruitful discussions and useful
comments.


\begin{thebibliography}{99}
%\addcontentsline{toc}{section}{References}

\bibitem{BizBM}
Bizyaev, I.\,A., Borisov, A.\,V., and Mamaev, I.\,S.,
The Hess\,--\,Appelrot System and~Its Nonholonomic Analogs,
\textit{Proc. Steklov Inst. Math.,}
2016, vol.\,294, pp.\,252--275;
see also:
\textit{Tr. Mat. Inst. Steklova,}
2016, vol.\,294, pp.\,268--292.


\bibitem{b21-75}
Borisov, A.\,V. and Mamaev, I.\,S.,
Strange Attractors in Rattleback Dynamics,
\textit{Physics–Uspekhi,}
2003, vol.\,46, no.\,4, pp.\,393--403;
see also:
\textit{Uspekhi Fiz. Nauk,}
2003, vol.\,173, no.\,4, pp.\,407--418.


\bibitem{KilinVed}
Kilin, A.\,A. and Vetchanin, E.\,V.,
The Control of~the~Motion through an~Ideal Fluid of~a~Rigid Body by~Means of~Two Moving Masses,
\textit{Nelin. Dinam.,}
2015, vol.\,11, no.\,4, pp.\,633--645
(Russian).


\bibitem{Ramod2002}
Kozlov, V.\,V. and Ramodanov, S.\,M.,
On the Motion of~a~Body with a~Rigid Hull and Changing Geometry of~Masses in an~Ideal Fluid,
\textit{Dokl. Phys.,}
2002, vol.\,47, no.\,2, pp.\,132--135;
see also:
\textit{Dokl. Akad. Nauk,}
2002, vol.\,382, no.\,4, pp.\,478--481.


\bibitem{Koz2016}
Kozlov, V.\,V.,
Invariant Measures of~Smooth Dynamical Systems, Generalized Functions and~Summation Methods,
\textit{Russian Acad. Sci. Izv. Math.,}
2016, vol.\,80, no.\,2, pp.\,342--358;
see also:
\textit{Izv. Ross. Akad. Nauk. Ser. Mat.,}
2016, vol.\,80, no.\,2, pp.\,63--80.


\bibitem{KozlovOn}
Kozlov, V.\,V. and Onishchenko, D.\,A.,
The Motion in~a~Perfect Fluid of~a~Body Containing a~Moving Point Mass,
\textit{J.~Appl. Math. Mech.,}
2003, vol.\,67, no.\,4, pp.\,553--564;
see also:
\textit{Prikl. Mat. Mekh.,}
2003, vol.\,67, no.\,4, pp.\,620--633.


\bibitem{KozlovRam}
Kozlov, V.\,V. and Ramodanov, S.\,M.,
The Motion of~a~Variable Body in~an~Ideal Fluid,
\textit{J.~Appl. Math. Mech.,}
2001, vol.\,65, no.\,4, pp.\,579--587;
see also:
\textit{Prikl. Mat. Mekh.,}
2001, vol.\,65, no.\,4, pp.\,592--601.


\bibitem{KozlovPMM2004}
Kozlov, V.\,V.,
Dynamics of~Variable Systems and~Lie Groups,
\textit{J.~Appl. Math. Mech.,}
2004, vol.\,68, no.\,6, pp.\,803--808;
see also:
\textit{Prikl. Mat. Mekh.,}
2004, vol.\,68, no.\,6, pp.\,899--905.


\bibitem{bizyaev_b21-3}% = chap
Chaplygin, S.\,A.,
On the Theory of Motion of Nonholonomic Systems. The Reducing-Multiplier
Theorem,
\textit{Regul. Chaotic Dyn.,}
2008, vol.\,13, no.\,4, pp.\,369--376;
see also:
\textit{Mat. Sb.,}
1912, vol.\,28, no.\,2, pp.\,303--314.


\bibitem{Fuf}
Fufaev, N.\,A.,
On the Possibility of~Realizing a~Nonholonomic Constraint by~Means of~Viscous Friction Forces,
\textit{J.~Appl. Math. Mech.,}
1964, vol.\,28, no.\,3, pp.\,630--632;
see also:
\textit{Prikl. Mat. Mekh.,}
1964, vol.\,28, no.\,3, pp.\,513--515.


\bibitem{AKN}
Arnol'd, V.\,I., Kozlov, V.\,V., and  Ne\u{\i}shtadt, A.\,I.,
\textit{Mathematical Aspects of Classical and Celestial Mechanics,}
3rd~ed.,
Encyclopaedia Math. Sci., vol.\,3,
Berlin: Springer, 2006.
%xiv+518 pp.


\bibitem{bizyaev}
Bizyaev, I.\,A.,
The Inertial Motion of~a~Roller Racer,
\textit{Regul. Chaotic Dyn.,}
2017, vol.\,22, no.\,3, pp.\,239--247.


\bibitem{bizyaevCylinder}
Bizyaev, I.\,A., Borisov, A.\,V., and Mamaev, I.\,S.,
Dynamics of~the~Chaplygin Sleigh on~a~Cylinder,
\textit{Regul. Chaotic Dyn.,}
2016, vol.\,21, no.\,1, pp.\,136--146.


\bibitem{Bolotin}
Bolotin, S. and Treschev, D.,
Unbounded Growth of Energy in Nonautonomous Hamiltonian Systems,
\textit{Nonlinearity,}
1999, vol.\,12, no.\,2, pp.\,365--388.


\bibitem{JacobiIntegral}
Borisov, A.\,V., Mamaev, I.\,S., and Bizyaev,~I.\,A.,
The Jacobi Integral in Nonholonomic Mechanics,
\textit{Regul. Chaotic Dyn.,}
2015, vol.\,20, no.\,3, pp.\,383--400.


\bibitem{HadamardHamel}
Borisov, A.\,V., Kilin, A.\,A., and~Mamaev,~I.\,S.,
On~the~Hadamard\,--\,Hamel Problem and the Dynamics of~Wheeled Vehicles,
\textit{Regul. Chaotic Dyn.,}
2015, vol.\,20, no.\,6, pp.\,752--766.


\bibitem{BM}
Borisov, A.\,V. and Mamaev, I.\,S.,
An~Inhomogeneous Chaplygin Sleigh,
\textit{Regul. Chaotic Dyn.,}
2017, vol.\,22, no.\,4, pp.\,435--447.


\bibitem{BM2}
Borisov, A.\,V. and Mamaev, I.\,S.,
The Dynamics of~a~Chaplygin Sleigh,
\textit{J.~Appl. Math. Mech.,}
2009, vol.\,73, no.\,2, pp.\,156--161;
see also:
\textit{Prikl. Mat. Mekh.,}
2009, vol.\,73, no.\,2, pp.\,219--225.


\bibitem{BorisovKuznetsov}
Borisov, A.\,V. and Kuznetsov, S.\,P.,
Regular and~Chaotic Motions of~Chaplygin Sleigh under Periodic Pulsed Torque Impacts,
\textit{Regul. Chaotic Dyn.,}
2016, vol.\,21, nos.\,7--8, pp.\,792--803.


\bibitem{BKM}
Borisov, A.\,V., Kilin, A.\,A., and Mamaev, I.\,S.,
The Problem of Drift and Recurrence for the Rolling Chaplygin Ball,
\textit{Regul. Chaotic Dyn.,}
2013, vol.\,18, no.\,6, pp.\,832--859.


\bibitem{ControlI}
Borisov, A.\,V., Kilin, A.\,A., and Mamaev, I.\,S.,
How to Control Chaplygin’s Sphere Using Rotors,
\textit{Regul. Chaotic Dyn.,}
2012, vol.\,17, nos.\,3--4, pp.\,258--272.


\bibitem{ControlII}
Borisov, A.\,V., Kilin, A.\,A., and Mamaev, I.\,S.,
How to Control Chaplygin’s Sphere Using Rotors:~2,
\textit{Regul. Chaotic Dyn.,}
2013, vol.\,18, nos.\,1--2, pp.\,144--158.


\bibitem{BorisovTop}
Borisov, A.\,V., Kazakov, A.\,O., and Sataev,~I.\,R.,
The Reversal and Chaotic Attractor in the Nonholonomic Model of Chaplygin's Top,
\textit{Regul. Chaotic Dyn.,}
2014, vol.\,19, no.\,6, pp.\,718--733.


\bibitem{Kuznetsov2012}
Borisov, A.\,V., Jalnine, A.\,Yu., Kuznetsov, S.\,P.,
Sataev,~I.\,R., and~Sedova,~J.\,V.,
Dynamical Phenomena Occurring due to Phase Volume Compression in Nonholonomic Model of the Rattleback,
\textit{Regul. Chaotic Dyn.,}
2012, vol.\,17, no.\,6, pp.\,512--532.


\bibitem{Caratheodory}
Carath\'{e}odory, C.,
Der Schlitten,
\textit{Z.~Angew. Math. Mech.,}
1933, vol.\,13, no.\,2, pp.\,71--76.


\bibitem{Fedorov}
Fedorov, Yu.\,N. and Garc{\'{\i}}a-Naranjo, L.\,C.,
The Hydrodynamic Chaplygin Sleigh,
\textit{J.~Phys.~A,}
2010, vol.\,43, no.\,43, 434013, 18\,pp.


\bibitem{Gonchenko2013}
Gonchenko, A.\,S., Gonchenko, S.\,V., and Kazakov, A.\,O.,
Richness of Chaotic Dynamics in Nonholonomic Models of~a~Celtic Stone,
\textit{Regul. Chaotic Dyn.,}
2013, vol.\,18, no.\,5, pp.\,521--538.


\bibitem{Turaev}
Gelfreich, V. and Turaev, D.,
Fermi Acceleration in Non-Autonomous Billiards,
\textit{J.~Phys.~A,}
2008, vol.\,41, no.\,21, 212003, 6\,pp.


\bibitem{Ito}
Ito, A.,
Successive Subharmonic Bifurcations and~Chaos in~a~Nonlinear Mathieu Equation,
\textit{Progr. Theoret. Phys.,}
1979, vol.\,61, no.\,3, pp.\,815--824.


\bibitem{Izrailev}
Izrailev, F.\,M., Rabinovich, M.\,I., and Ugodnikov, A.\,D.,
Approximate Description of~Three-Dimensional Dissipative Systems with~Stochastic Behaviour,
\textit{Phys. Lett.~A,}
1981, vol.\,86, nos.\,6--7, pp.\,321--325.


\bibitem{Pereira}
Pereira, T. and Turaev, D.,
Exponential Energy Growth in~Adiabatically Changing Hamiltonian Systems,
\textit{Phys. Rev.~E~(3),}
2015, vol.\,91, no.\,1, 010910(R), 4\,pp.


\bibitem{Jung}
Jung, P., Marchegiani, G., and Marchesoni,~F.,
Nonholonomic Diffusion of~a~Stochastic Sled,
\textit{Phys. Rev.~E,}
2016, vol.\,93, no.\,1, 012606, 9\,pp.


\bibitem{Krishnaprasad}
Krishnaprasad, P.\,S. and Tsakiris, D.\,P.,
Oscillations, ${\rm SE}(2)$-Snakes and Motion Control: A~Study of the Roller Racer,
\textit{Dyn. Syst.,}
2001, vol.\,16, no.\,4, pp.\,347--397.


\bibitem{KellyBeanie}
Kelly, S.\,D., Fairchild, M.\,J., Hassing, P.\,M., and~Tallapragada,~P.,
Proportional Heading Control for~Planar Navigation: The~Chaplygin Beanie and~Fishlike Robotic Swimming,
in
\textit{Proc. of~the~American Control Conf. (Montreal,\,QC, Canada, June~2012),} pp.\,4885--4890.


\bibitem{Koiller}
Koiller, J., Markarian, R., Oliffson Kamphorst, S., and~Pinto de~Carvalho,~S.,
Time-Dependent Billiards,
\textit{Nonlinearity,}
1995, vol.\,8, no.\,6, pp.\,983--1003.


\bibitem{Pivovarova}
Kilin, A.\,A., Pivovarova, E.\,N., and~Ivanova,~T.\,B.,
Spherical Robot of Combined Type: Dynamics and Control,
\textit{Regul. Chaotic Dyn.,}
2015, vol.\,20, no.\,6, pp.\,716--728.


\bibitem{Ott}
Ott, E., Grebogi, C., and~Yorke,~J.\,A.,
Controlling Chaos,
\textit{Phys. Rev. Lett.,}
1990, vol.\,64, no.\,11, pp.\,1196--1199.


\bibitem{Leonard}
Leonard, N.\,E.,
Periodic Forcing, Dynamics and~Control of~Underactuated Spacecraft and~Underwater Vehicles,
in
\textit{Proc. of~the~34th~IEEE Conf. on~Decision and~Control (New Orleans,\,La., Dec~1995),} pp.\,3980--3985.


\bibitem{Lenz}
Lenz, F., Diakonos, F.\,K., and~Schmelcher,~P.,
Tunable Fermi Acceleration in the Driven Elliptical Billiard,
\textit{Phys. Rev. Lett.,}
2008, vol.\,100, no.\,1, 014103, 4\,pp.


\bibitem{Ostrowskiy}
Lewis, A.\,D., Ostrowskiy, J.\,P., Burdickz, J.\,W., and~Murray,~R.\,M.,
Nonholonomic Mechanics and~Locomotion: The~Snakeboard Example,
in
\textit{Proc. of~the~IEEE Internat. Conf. on~Robotics and~Automation (San Diego,\,Calif., May~1994),}
pp.\,2391--2400.


\bibitem{Liouville}
Liouville, J.,
D\'eveloppements sur un~chapitre de~la~M\'ecanique de~Poisson,
\textit{J.~Math. Pures Appl.,}
1858, vol.\,3, pp.\,1--25.


\bibitem{Lichtenberg}
Lichtenberg, A.\,J. and Lieberman, M.\,A.,
\textit{Regular and~Chaotic Dynamics,}
2nd~ed.
Appl. Math. Sci., vol.\,38,
New York: Springer, 1992.
%692\,pp.


\bibitem{Murray}
Murray, R.\,M. and Sastry, S.\,Sh.,
Nonholonomic Motion Planning: Steering Using Sinusoids,
\textit{IEEE Trans. Automat. Control,}
1993, vol.\,38, no.\,5, pp.\,700--716.


\bibitem{Zenkov}
Osborne, J.\,M. and Zenkov, D.\,V.,
Steering the~Chaplygin Sleigh by~a~Moving Mass,
in
\textit{Proc. of~the~44th IEEE Conf. on~Decision and~Control (Seville, Spain, Dec~2005),} pp.\,1114--1118.


\bibitem{Sprott}
Sprott, J.\,C.,
\textit{Elegant Chaos: Algebraically Simple Chaotic Flows,}
Singapore: World Sci., 2010.
%304pp


\bibitem{Tallapragada}
Tallapragada, P. and Kelly, S.\,D.,
Integrability of~Velocity Constraints Modeling Vortex Shedding in~Ideal Fluids,
\textit{J.~Comput. Nonlinear Dynam.,}
2017, vol.\,12, no.\,2, 021008, 7\,pp.


\bibitem{KellyAbrajan}
Kelly, S.\,D. and Abrajan-Guerrero, R.,
Planar Motion Control, Coordination, and~Dynamic Entrainment for~a~Singly Actuated Nonholonomic Robot,
\textsf{http://scottdavidkelly.wdfiles.com/local--files/start/kellyabrajan-guerrero16cdc.pdf}
(2016).
%6pp


\bibitem{Vetchanin}
Vetchanin, E.\,V. and Kilin, A.\,A.,
Free and Controlled Motion of~a~Body with Moving Internal Mass though a~Fluid in the Presence
of Circulation around the Body,
\textit{Dokl. Phys.,}
2016, vol.\,61, no.\,1, pp.\,32--36;
see also:
\textit{Dokl. Akad. Nauk,}
2016, vol.\,466, no.\,3, pp.\,293--297.


\bibitem{JungScatt}
Jung, Ch. and Scholz, H.-J.,
Chaotic Scattering off~the~Magnetic Dipole,
\textit{J.~Phys.~A,}
1988, vol.\,21, no.\,10, pp.\,2301--2311.


\bibitem{Eckhardt}
Eckhardt, B. and Jung, C.,
Regular and~Irregular Potential Scattering,
\textit{J.~Phys.~A,}
1986, vol.\,19, no.\,14, L829--L833.


\bibitem{Aref}
Toph{\o}j, L. and Aref, H.,
Chaotic Scattering of~Two Identical Point Vortex Pairs Revisited,
\textit{Phys. Fluids,}
2008, vol.\,20, 093605, 10\,pp.

\bibitem{Kuznetsov}
Bizyaev, I.\,A., Borisov, A.\,V., and Kuznetsov, S.\,P.,
Chaplygin Sleigh with~Periodically Oscillating Internal Mass,
\textit{Europhys. Lett.,}
2017, vol.\,119, no.\,6, 60008, 7\,pp.

\bibitem{Biz-A1}
Feller, W.,
\textit{An Introduction to Probability Theory and its Applications,}
3rd~ed.,
vol.\,1,
New York: Wiley, 1968.

\bibitem{Biz-A2}
Rytov, S.\,M., Kravtsov, Y.\,A., Tatarskii, V.\,I.,
\textit{Principles of Statistical Radiophysics. 1. Elements of Random Process Theory,}
Berlin: Springer, 1987.

\bibitem{Biz-A3}
Cox, D.\,R., Miller, H.\,D.,
\textit{The Theory of Stochastic Processes,}
New York: Chapman and Hall/CRC, 2017.

\bibitem{Biz-A4}
Borisov, A.\,V., Mamaev, I.\,S., Bizyaev, I.\,A.,
Dynamical systems with non-integrable constraints: vaconomic mechanics, sub-Riemannian geometry, and non-holonomic mechanics,
\textit{Uspekhi Mat. Nauk,}
2017, vol.\,72, no.\,5(437), pp.\,3--62.

\bibitem{Biz-A5}
Kuznetsov, S.\,P.,
Plate Falling in a Fluid: Regular and Chaotic Dynamics of Finite-dimensional Models,
\textit{Regul. Chaotic Dyn.,}
2015, vol.\,20, no.\,3, pp.\,345--382.

\bibitem{45}
Bizyaev, I.\,A., Borisov, A.\,V., and Mamaev, I.\,S.,
The Dynamics of Nonholonomic Systems Consisting of~a~Spherical Shell with~a~Moving Rigid Body Inside,
\textit{Regul. Chaotic Dyn.,}
2014, vol.\,19, no.\,2, pp.\,198--213.

\end{thebibliography}
\end{document}